\documentclass[aps,prstab, superscriptaddress]{revtex4-1}

\usepackage{graphicx} 

\begin{document}


\title{{\huge The Next SRF Technologies}\\
\vspace{0.5cm}
\textsf{Talk presented at the International Workshop on Future Linear Colliders (LCWS2016), \\
Morioka, Japan, 5-9 December 2016. 
C16-12-05.4.}}



\author{\large Takayuki Kubo}
\email[]{kubotaka@post.kek.jp}
\affiliation{KEK (High Energy Accelerator Research Organization), Tsukuba, Ibaraki 305-0801, Japan.}
\affiliation{SOKENDAI (the Graduate University for Advanced Studies), Hayama, Kanagawa 240-0115, Japan.}



\begin{abstract}
In this talk, I introduce the proposed next superconducting radio-frequency (SRF) technologies that will make it possible to achieve much higher accelerating electric field than the present SRF technologies. 
Audiences are assumed to be non-experts. 
We start from a brief review of basics of SRF, 
history of the high gradient technologies and the layered structure behind it. 
The multiple benefit of the layered structure is introduced. 
We then move to the next SRF technologies: 
superconductor-superconductor (SS) structure and superconductor-insulator-superconductor (SIS) structure. 
We discuss the SS structure in detail. 
Experimental results are also introduced and compared with theoretical considerations. 
\end{abstract}

%
%


\maketitle


\section{Introduction}

%
\begin{figure}[t]
   \begin{center}
   \includegraphics[width=1\linewidth]{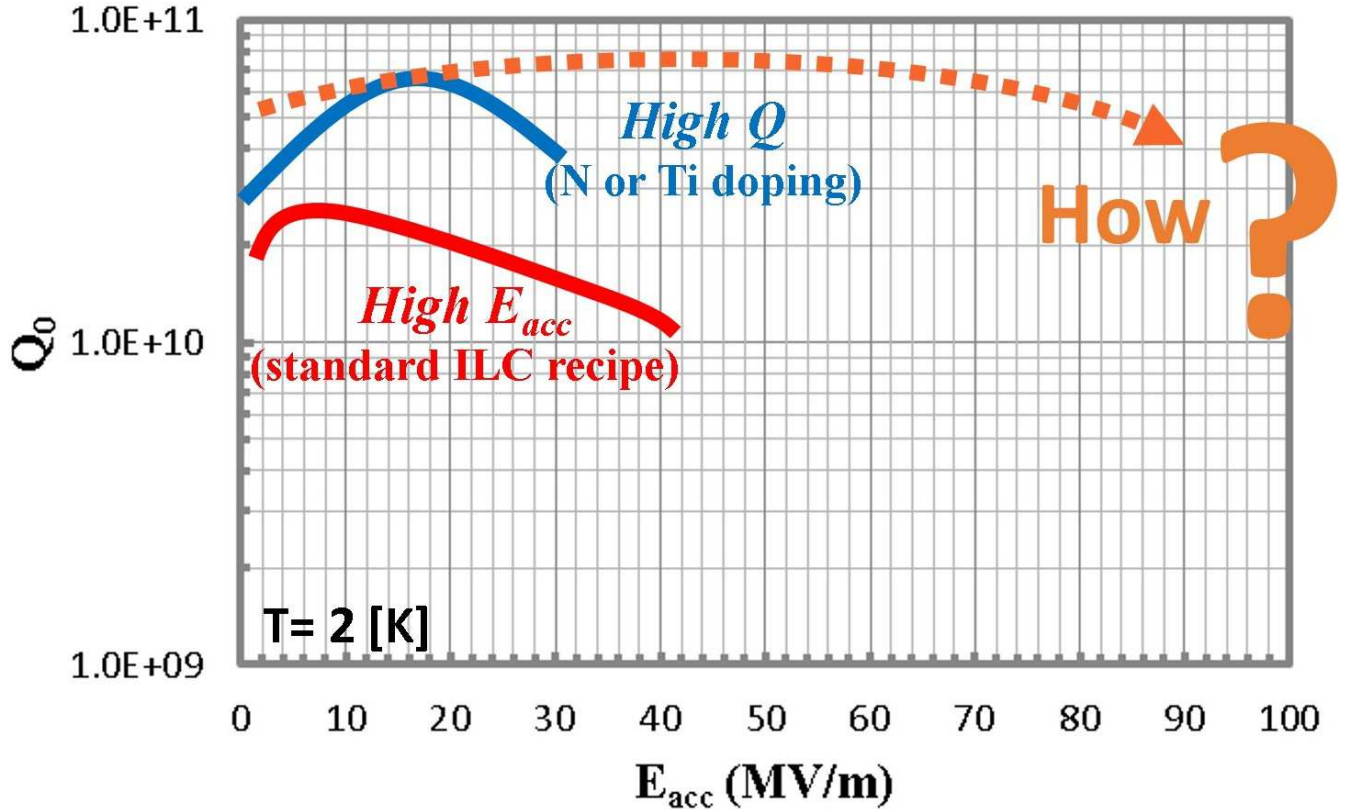}
   \end{center}\vspace{0cm}
   \caption{
Rough sketches of typical performances of Nb cavities processed by the present SRF technologies: the standard International Linear Collider recipe (red curve) and the impurity doping (blue curve). 
The orange dotted curve represents our dream for the future. 
   }\label{fig1}
\end{figure}

Superconducting radio-frequency (SRF) cavities accelerate electrons by using the electric field parallel to the axis. 
The average electric field that the charged particle sees during transit is called the accelerating field $E_{acc}$. 
As $E_{acc}$ increases, 
the necessary accelerator length to achieve a target particle energy decreases. 
At the same time, we want to reduce dissipation at the inner surface of the cavity: 
a small surface resistance $R_s$ or a large quality factor $Q_0 \propto 1/R_s$ is necessary. 
Thus these two parameters, $E_{acc}$ and $Q_0$, are used to describe performances of SRF cavities, and performance test results are plotted on $Q_0$-$E_{acc}$ planes.

The red curve in Fig.~\ref{fig1} represents the rough sketch of the performance of a typical Nb cavity processed by the standard International Linear Collider (ILC) recipe. 
As $E_{acc}$ increases, $Q_0$ increases a little and then continues to decline. 
The maximum field is given by the right endpoint of the curve, $E_{acc}\simeq 40\,{\rm MV/m}$, 
where $Q_0 > 10^{10}$. 
This is the high gradient technology established in 1990s. 
We have another technology established in 2010s, 
the so-called high-Q technology~\cite{Grassellino_N_dope, Dhakal_Ti_dope_PRAB, Dhakal_Ti_dope_IPAC2012}. 
It can drastically improve $Q_0$, 
but the maximum value of $E_{acc}$ is reduced. 
See the blue curve in Fig.~\ref{fig1}. 
As $E_{acc}$ increases, $Q_0$ significantly increases, then start to decrease. 
The quench field is rather low compared with that of ILC recipe. 
The present SRF technologies consist of these two technologies: 
high-gradient and high-$Q$ technologies. 
Considering cavity performances in pioneer days of SRF history (1960s), 
progresses in the past several decades are truly remarkable. 
However, we are not satisfied with the present technologies. 
We want to go beyond the limits of the present technologies and to realize the dotted orange curve. 
But, how can we do that? 
That's the topic I want to talk about today. 
Since participants are not necessarily experts of SRF, 
we begin with basics of SRF focusing on high gradient applications.

\section{Present high-gradient technology}

\subsection{Basics of SRF for higher gradients} \label{section:basics}

%
\begin{figure*}[tb]
   \begin{center}
   \includegraphics[width=0.95\linewidth]{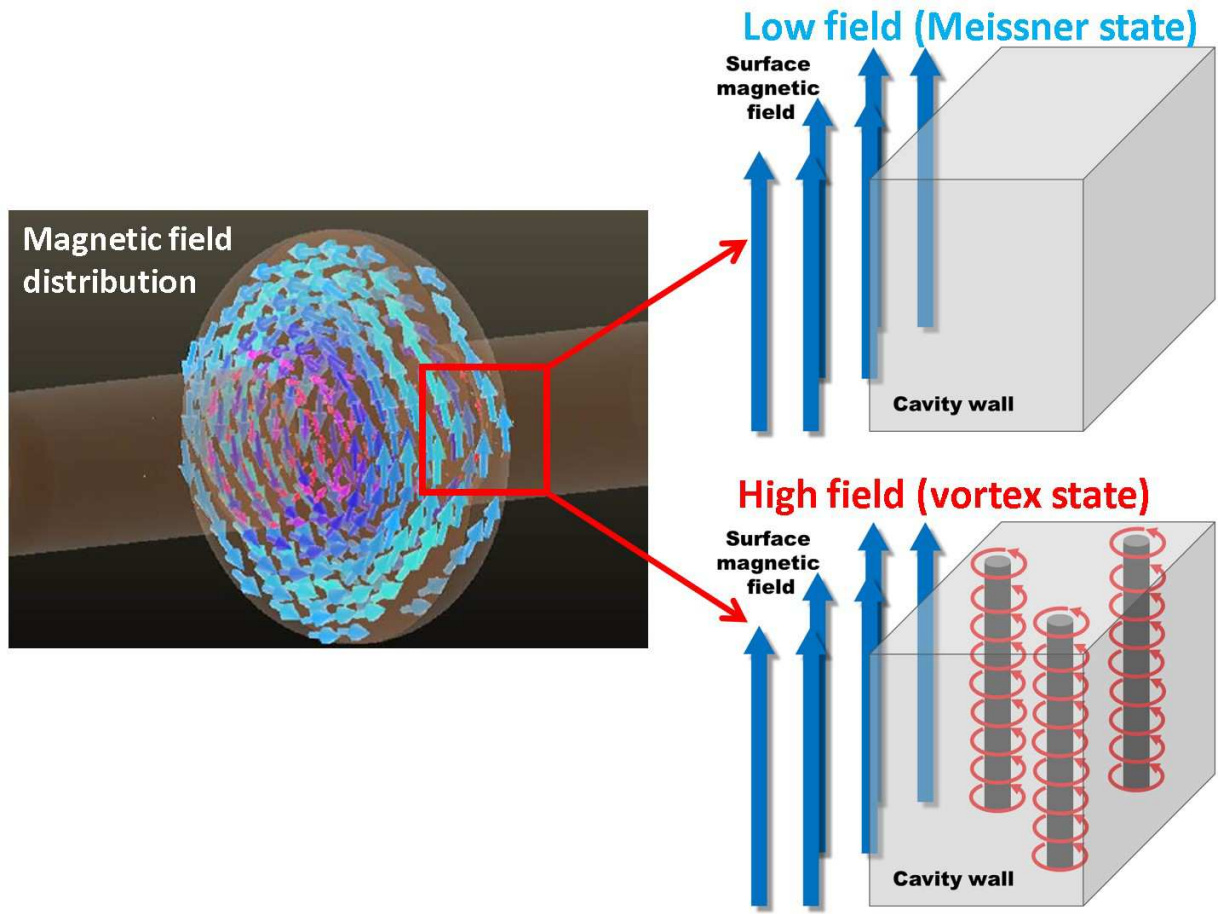}
   \end{center}\vspace{0cm}
   \caption{
The magnetic field distribution in a cavity. 
The upper right figure represents a sketch of the cavity wall at a low enough magnetic field, 
where the superconducting wall expels the magnetic flux (Meissner state). 
On the other hand, the lower right figure represents the cavity wall at a high magnetic field, 
where vortices penetrate into the superconducting wall (vortex state).  
   }\label{fig2}
\end{figure*}

Inside the SRF cavity, the magnetic field distributes like the left figure in Fig.~\ref{fig2}. 
When $E_{acc}$ is small, the surface magnetic field $B_0$ is also small, 
and the magnetic flux is expelled to the outside the superconductor by the Meissner effect (see the upper right figure in Fig.~\ref{fig2}). 
However, as $E_{acc}$ increases, $B_0$ also increases, 
and finally vortices start to penetrate into the inside of the superconducting cavity wall (see the lower right figure in Fig.~\ref{fig2}). 
This state is called the vortex state. 
The vortex is a normal conducting filament oscillating with the RF frequency ($\sim \,{\rm GHz}$). 
Dissipation due to the vortex dynamics is so huge and can lead to quenches: 
the vortex state cannot be used for operations of SRF cavities. 
To achieve a high $E_{acc}$ (or high $B_0$), 
a material that can withstand against the vortex penetration up to a high magnetic field should be used. 
That is one of the essential reasons we are using Nb as the material of SRF cavity. 
The lower critical field of pure Nb is $B_{c1}\simeq 170\,{\rm mT}$ ($E_{\rm acc}\simeq 40\,{\rm MV/m}$ for TESLA cavity), 
which is larger than other superconductors. 
(It should be noted that another important reason to use pure material comes from its thermal conductivity: 
as its purity increases, the thermal conductivity improves and stabilize cavities against the thermal breakdown.)

Now we know we should use pure Nb. 
But achieving such a high field $B_0 \sim B_{c1}\simeq 170\,{\rm mT}$ is not a straight forward task, 
even if ultra pure Nb is available. 
Next we will look back the history of SRF and review how SRF researchers achieved $B_0 > 170\,{\rm mT}$ ($E_{\rm acc} > 40\,{\rm MV/m}$). 
The history would provide us with a clue to a higher $E_{\rm acc}$.

\subsection{History and finding of low temperature baking}\label{section:history}

%
\begin{figure*}[tb]
   \begin{center}
   \includegraphics[width=1\linewidth]{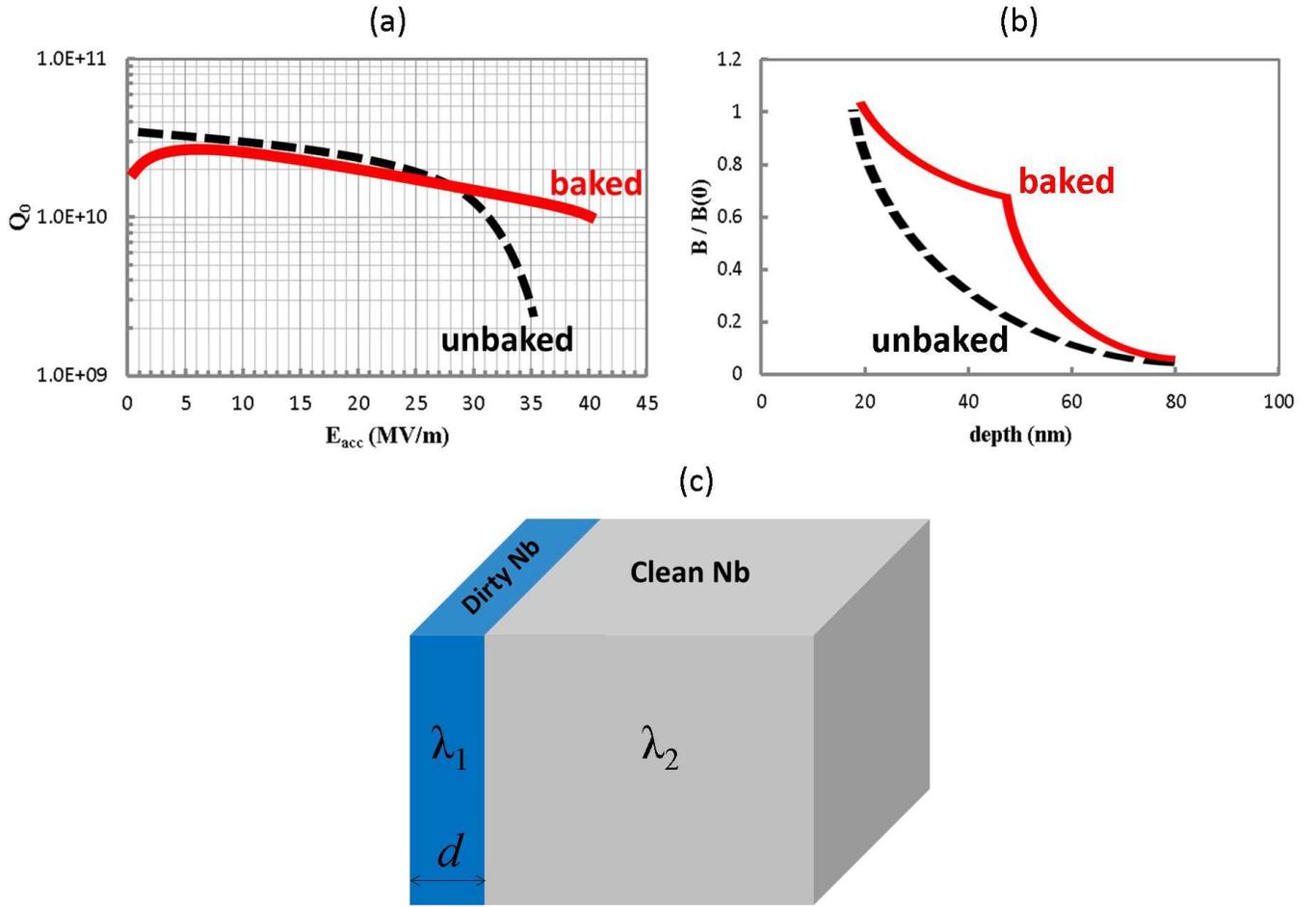}
   \end{center}\vspace{0cm}
   \caption{
(a) Rough sketch of typical $Q$-$E$ curves for baked (solid red curve) and unbaked cavities (dashed black curve). 
Real data is found in Ref.~\cite{lilje_bake}, for example. 
(b) Rough sketch of the magnetic field distribution in baked (solid red curve) and unbaked Nb (dashed black curve). 
Real data obtained by using the low energy muon spin rotation is found in Ref.~\cite{Romanenko_bake}. 
   }\label{fig3}
\end{figure*}

When we look back the history of SRF (see figures in Ref.~\cite{Ciovati_IPAC2013}), 
we realize the record field had increased year by year. 
Various ideas had pushed up the record field:
electropolishing (EP) followed by a heat treatment for hydrogen degassing~\cite{saitoEP, furuyaEP}, high-pressure rinsing~\cite{bernerdHPR, saitoHPR, kneiselHPR}, clean assembly~\cite{kojimaCA} etc. 
The milestone in the high-gradient quest was the achievement of $B_0\simeq B_{c1}$ or $E_{\rm acc}\simeq 40\,{\rm MV/m}$ in 1990s (the record field has changed little up to the present date since 1990s).
The key was the low temperature baking~\cite{kako_bake, kneisel_40, ono_bake, lilje_bake}. 
Fig.~\ref{fig3}(a) shows rough sketches of performances of Nb cavities with and without the low temperature baking after EP. 
The $Q$ value of unbaked cavity (dashed black curve) suddenly drops at $E_{\rm acc}\simeq 30\,{\rm MV/m}$: the so-called high field $Q$-drop. 
On the other hand, the baked cavity (solid red curve) does not show such a rapid Q decay at a high field and can achieve $E_{\rm acc}\simeq 40\,{\rm MV/m}$.

Today we know the baked Nb has the layered structure that consists of a thin dirty Nb layer and bulk clean Nb. 
This fact was clearly shown in the beautiful experiment using the low energy muon spin
rotation (LE-$\mu$SR)~\cite{Romanenko_bake}. 
Let us see the rough sketch shown in Fig.~\ref{fig3}(b). 
In the unbaked Nb, the magnetic field rapidly decay as we go into the inside (see the dashed black curve). 
On the other hand, in the baked Nb, 
the magnetic field attenuates slower than the unbaked Nb in the first $50$-$60\,{\rm nm}$ and then rapidly decay at a deeper region (see the solid red curve). 
This means that the surface is dirty and the inside is clean. 
Then baked Nb can be modeled like Fig.~\ref{fig3}(c): 
the layered structure that consists of the thin dirty Nb with thickness $d$ and penetration depth $\lambda_1$ and bulk clean Nb with $\lambda_2\, (< \lambda_1)$.

\subsection{Multiple benefit of the layered structure}\label{section:benefit}

In the present day, we know many properties of the layered Nb structure through a number of theoretical studies so far. 
It would be beneficial to summarize them. 

\subsubsection{Well behaved density of states at a high field}\label{section:DOS}

%
\begin{figure*}[tb]
   \begin{center}
   \includegraphics[width=1\linewidth]{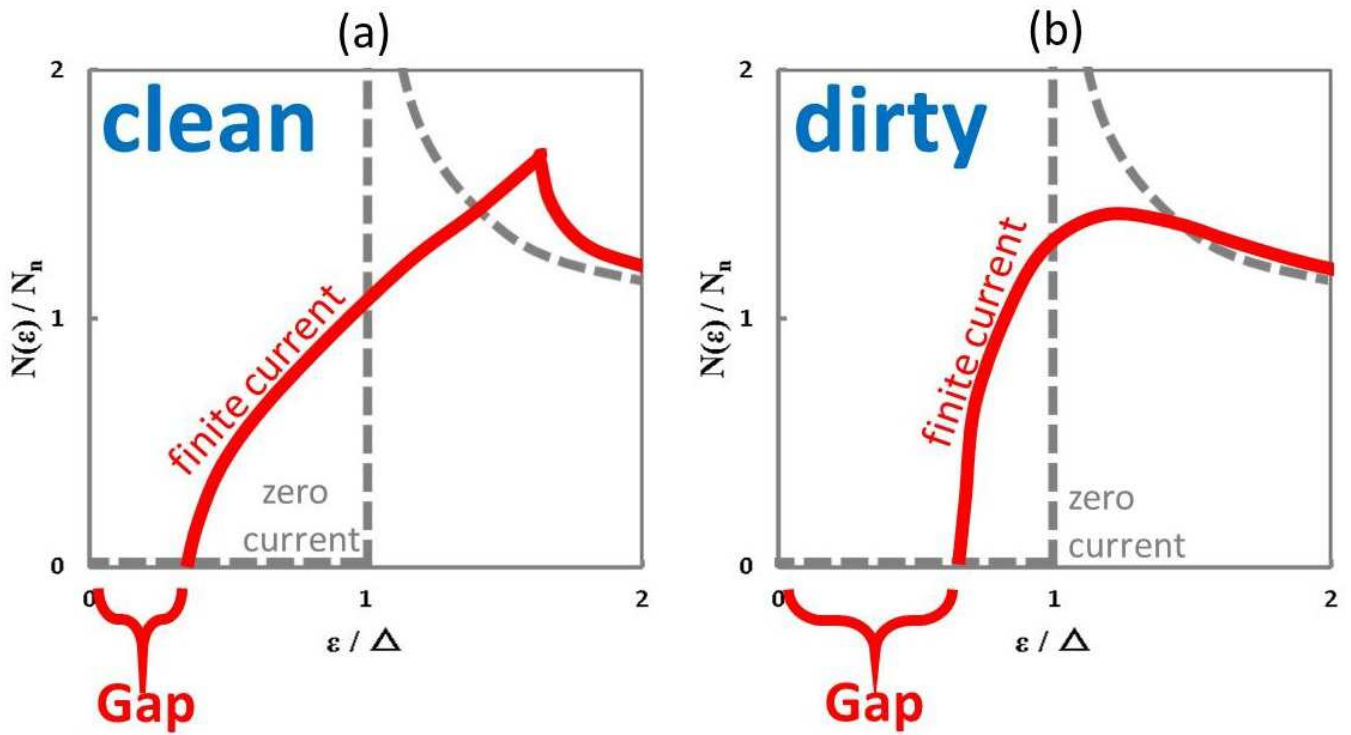}
   \end{center}\vspace{0cm}
   \caption{
Rough sketches of density of states for (a) clean and (b) dirty superconductors. 
The dashed gray curves represent density of states under the zero current. 
The solid red curves represent density of states under a finite current. 
Calculations and detailed discussions are found in Refs.~\cite{Lin_Gurevich, Gurevich_review}. 
   }\label{fig4}
\end{figure*}

The first benefit is the well behaved density of states (DOS) of the dirty Nb at a high field~\cite{Lin_Gurevich, Gurevich_review}. 
Fig.~\ref{fig4}(a) and (b) show rough sketches of the DOS of clean and dirty superconductors, respectively. 
When the surface current is zero, 
the DOS of the clean and dirty superconductor are given by the common shape (dashed gray curves). 
As the surface current increases, 
the DOS of the clean superconductor becomes different from that of the dirty one (solid red curves). 
In particular, the gap of the clean superconductor rapidly decreases and becomes gapless before arriving at the superheating field $B_s$, 
while the dirty superconductor still has a finite gap at $B_s$ (see the figures in Refs.~\cite{Lin_Gurevich, Gurevich_review}). 
The dirty superconductor is rather well behaved at a high field.

Use of a simple bulk dirty Nb, however, suffers a reduction of an achievable field due to a quench triggered by vortex penetration at $B_0 \sim B_{c1}^{\rm (dirty\,Nb)} < B_{c1}^{\rm (clean\,Nb)}\simeq 170\,{\rm mT}$ (remind that we have repeatedly observed the quenches of impurity doped cavities at $B_0 < 170\,{\rm mT}$). 
This difficulty is avoidable by using the layered structure that consists of a thin dirty Nb layer and a clean bulk Nb. 
This structure retains the $B_{c1}$ of clean Nb and avoids the significant reduction of gap, 
because $B_{c1}$ is a bulk property and given by $B_{c1} = B_{c1}^{\rm (clean\,Nb)}\simeq 170\,{\rm mT}$ and the gap reduction is suppressed at the surface dirty layer.

The high-field Q drop observed in the unbaked clean Nb is thought to be caused by the gap reduction due to the surface current, 
and the cure of high-field Q drop by the low temperature baking comes from the transformation of the surface thin layer from the clean one to dirty one~\cite{Lin_Gurevich, Gurevich_review}.

\subsubsection{Surface current suppression and field limit enhancement}\label{section:field_limit}

%
\begin{figure*}[tb]
   \begin{center}
   \includegraphics[width=0.6\linewidth]{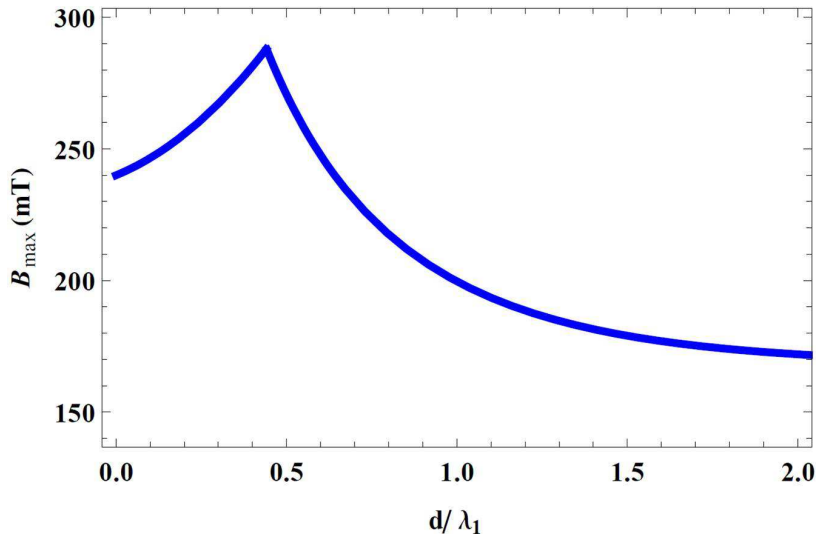}
   \end{center}\vspace{0cm}
   \caption{
Theoretical field limit of a dirty Nb coated clean Nb substrate. 
The horizontal axis represents the dirty Nb thickness $d$. 
Here $\lambda_1=180\,{\rm nm}$, $B_s^{\rm (dirty\,Nb)}=0.84 B_c^{\rm (Nb)} = 170\,{\rm mT}$, $\lambda_2=40\,{\rm nm}$, and $B_s^{\rm (clean\,Nb)}=240\,{\rm mT}$ are assumed. 
Note that this figure shows the theoretical field limit for the ideal case. 
For a realistic case, there exist surface defects, and the achievable field becomes much different from the figure. 
For example, consider the case $d \gtrsim \lambda_1$. 
This is just a bulk dirty Nb. 
Then the practically achievable field should be given  by $B_{c1}^{\rm (dirty\,Nb)}$ instead of $B_{s}^{\rm (dirty\,Nb)}$. 
In general, a ``gimmick" to stop vortex penetration is necessary to go beyond $B_{c1}$. 
Without a gimmick, the practically achievable field remains $B_0 \sim B_{c1}$ even if the theoretical field limit is enhanced. 
The existence of such a gimmick is the important aspect of the layered structure discussed in Section~\ref{section:additional_barrier}. 
   }\label{fig5}
\end{figure*}

When $\lambda_1 > \lambda_2$, 
the magnetic field attenuation in the surface layer is prevented by the counterflow due to the current in the substrate~\cite{Kubo_APL, Kubo_LINAC2014, Gurevich_AIP, Posen_PhysRev, Kubo_SUST}. 
As a result, the surface current is suppressed and the theoretical field limit is enhanced, 
because the theoretical field limit is determined by the current density. 
In addition, the suppression of the surface current further delays the gap reduction to a higher field.

Fig.~\ref{fig5} shows the theoretical field limit of the layered Nb system that consists of a surface dirty Nb layer and bulk clean Nb~\cite{Kubo_LINAC2014, Gurevich_AIP, Kubo_SUST}. 
It should be noted that this is the theoretical field limit for the ideal case and does not necessarily corresponds with the practically achievable field. 
Taking into account the existence of surface defects, 
the achievable field becomes much different. 
For example, consider the case $d \gg \lambda_1$, 
which is nothing but a bulk dirty Nb. 
In this case, the Meissner state ceases to be stable at $B_0 \sim B_{c1}^{\rm (dirty\,Nb)}$. 
While the Bean-Livingstone (BL) surface barrier prevent vortex entrance, 
it cannot adequately protect the superconductor, 
because topographic and material defects covering the surface weaken the barrier. 
Then the practically achievable field for this case is given  by $B_{c1}^{\rm (dirty\,Nb)}$ instead of $B_{s}^{\rm (dirty\,Nb)}$. 
In general, 
to go beyond $B> B_{c1}$, we need a ``gimmick" to stop vortex penetration. 
Otherwise, even if the theoretical field limit is enhanced, 
the practically achievable field remains $B_0 \sim B_{c1}$. 
As seen below, the existence of such a gimmick is the important aspect of the layered Nb structure.

\subsubsection{Additional barrier against vortex penetration}\label{section:additional_barrier}

%
\begin{figure*}[tb]
   \begin{center}
   \includegraphics[width=0.9\linewidth]{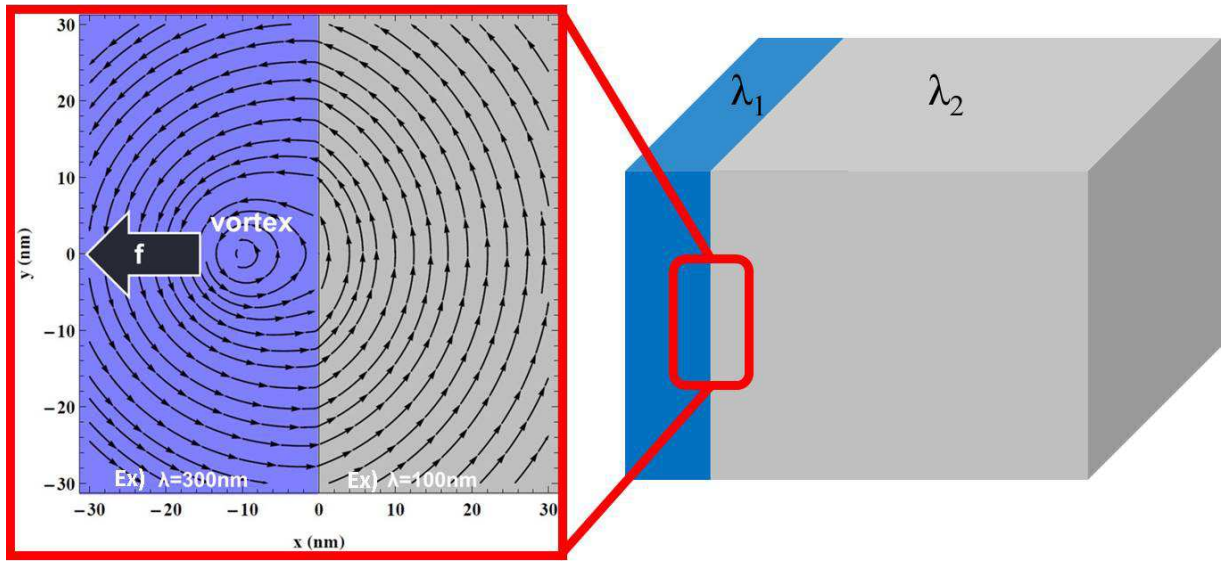}
   \end{center}\vspace{0cm}
   \caption{
A vortex located in the vicinity of the interface of two superconductors with different penetration depth. 
   }\label{fig6}
\end{figure*}

In the layered structure with $\lambda_1 > \lambda_2$, 
the boundary of two superconductors plays the role of the additional barrier~\cite{Kubo_LINAC2014}: 
a vortex at the region of $\lambda_1$ is expelled by the interface with the region of $\lambda_2$ (see Fig.~\ref{fig6}). 
This barrier originates from image vortices introduced to satisfy the boundary condition at the interface of two superconductors with different penetration depths. 
The major contribution comes from the nearest image with the same sign, 
which expels the vortex to the direction of the material with a larger penetration depth. 
The detailed discussion and derivation are found in Ref.~\cite{Kubo_SUST}. 
Note that the mechanism of the SS boundary barrier is different from that of the BL barrier: 
remind the BL barrier comes from the image with the opposite sign, 
which attracts the vortex to the surface. 
The discussion based on the Ginzburg-Landau theory is found in Ref.~\cite{Checchin}. 

\subsubsection{Surface resistance variation}\label{section:Rs}

%
\begin{figure*}[tb]
   \begin{center}
   \includegraphics[width=0.6\linewidth]{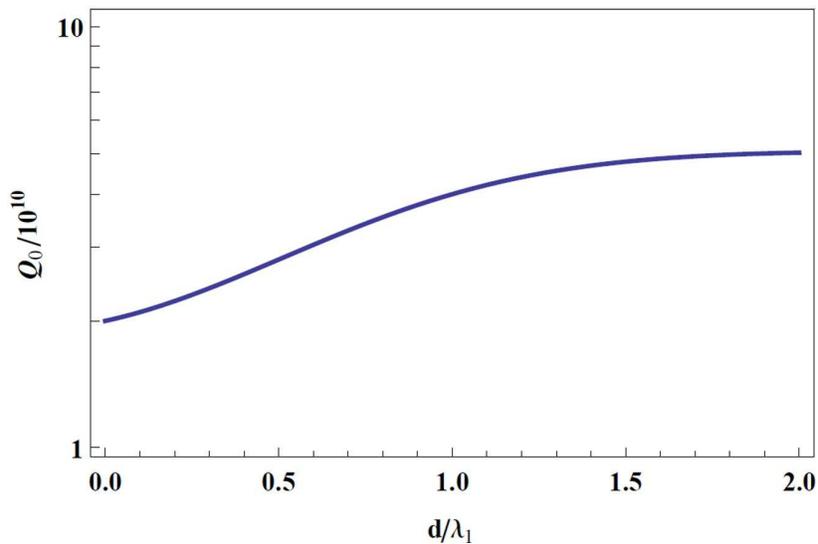}
   \end{center}\vspace{0cm}
   \caption{
$Q_0$ value of a layered Nb structure at $2\,{\rm K}$ as a function of the surface layer  thickness $d$, 
where $Q_0^{\rm (layer)}=5\times 10^{10}$ and $Q_0^{\rm (sub)}=2\times 10^{10}$ are assumed.  
   }\label{fig7}
\end{figure*}

Since the current flows on both the materials when the surface layer is thinner than the penetration depth ($d \lesssim \lambda_1$), 
the surface resistance is given by a linear combination, $R_s = a R_s^{\rm (layer)} + b R_s^{\rm (sub)}$, 
where $R_s^{\rm (layer)}$ and $R_s^{\rm (sub)}$ are the surface resistance of the layer material and the substrate material, respectively. 
The concrete expressions of the factors $a$ and $b$ are given in Refs.~\cite{Gurevich_AIP, Kubo_SUST}. 
In particular, when the surface layer material has a smaller surface resistance $R_s^{\rm (layer)} < R_s^{\rm (sub)}$, 
the total surface resistance $R_s$ can be smaller than $R_s^{\rm (sub)}$. 
This effect might be related to the relatively high Q results obtained by recently developped nitrogen infusion recipe~\cite{Grassellino_bake, Grassellino_bake_arxiv}. 
Fig.~\ref{fig7} shows an example of $Q_0$ value of a layered Nb for the case the surface layer has a smaller surface resistance than the bulk.

\section{What's next?}

\subsection{Advanced layered structures: SS and SIS}\label{section:next}

%
\begin{figure*}[tb]
   \begin{center}
   \includegraphics[width=0.75\linewidth]{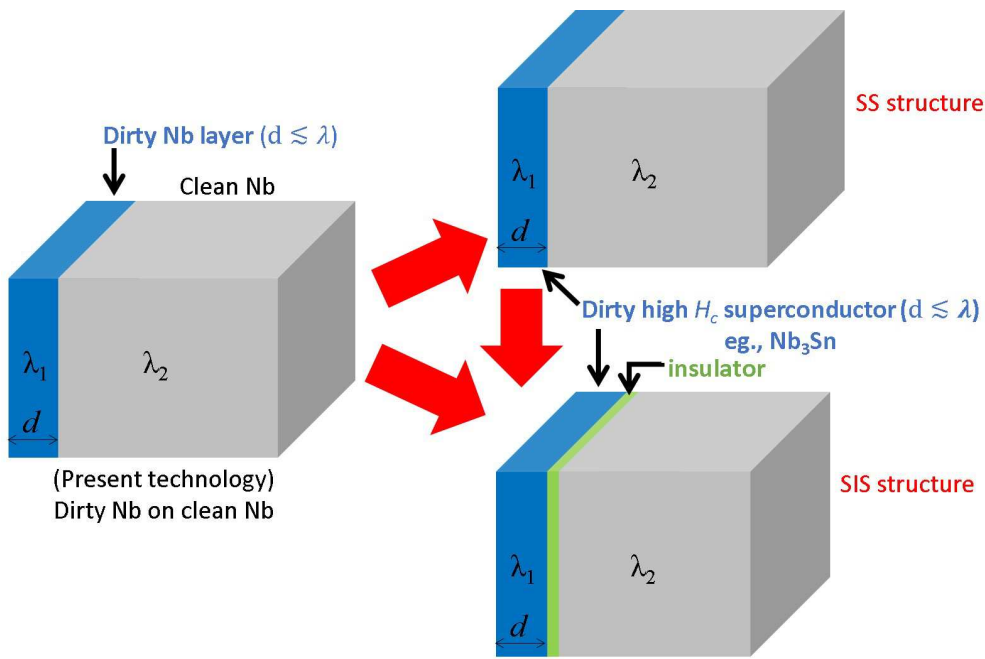}
   \end{center}\vspace{0cm}
   \caption{
Present and next SRF technologies. 
   }\label{fig8}
\end{figure*}

\begin{figure*}[tb]
   \begin{center}
   \includegraphics[width=1\linewidth]{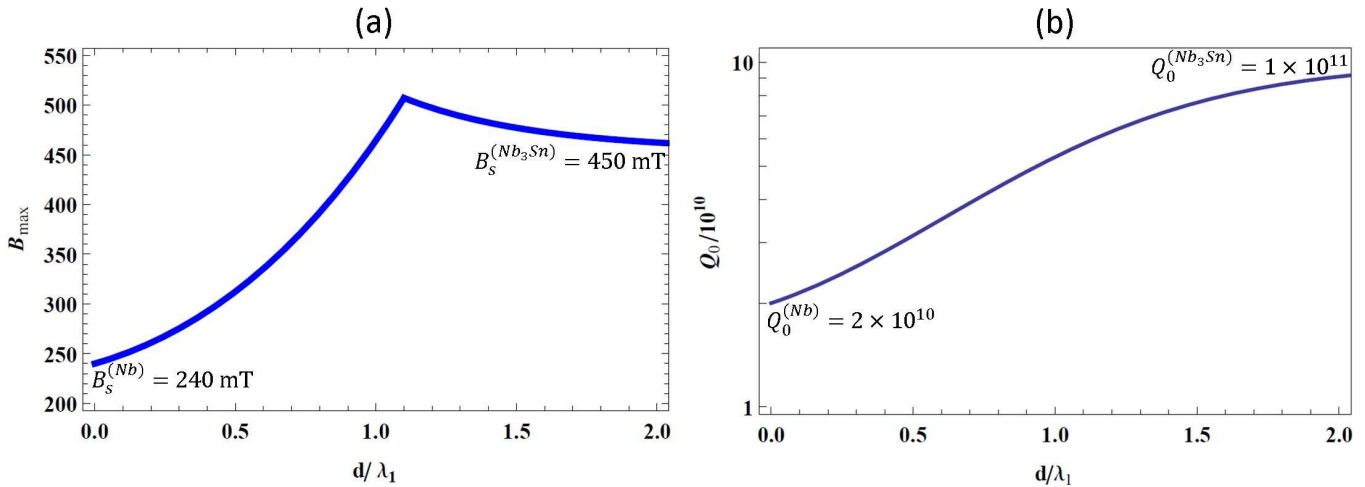}
   \end{center}\vspace{0cm}
   \caption{
(a) Theoretical field limit of SIS with a thin insulator or SS structure that consists of ${\rm Nb_3 Sn}$ layer and bulk Nb as a function of the ${\rm Nb_3 Sn}$ thickness $d$, 
where $\lambda_1=120\,{\rm nm}$, $B_s^{\rm (Nb_3 Sn)}=450\,{\rm mT}$, $\lambda_2=40\,{\rm nm}$, and $B_s^{\rm (Nb)}=240\,{\rm mT}$ are assumed. 
Note here this is a theoretical field limit for the ideal case. 
For a realistic case, there exist surface defects, and the practically achievable field becomes much different from the figure. 
For example, consider the case $d \gtrsim \lambda_1$, 
which is nothing but a bulk ${\rm Nb_3 Sn}$. 
Then the achievable field should be given by $B_0\sim B_{c1}^{\rm (Nb_3 Sn)}$ instead of $B_{s}^{\rm (Nb_3 Sn)}$. 
(b) $Q_0$ value at $2\,{\rm K}$ as a function of $d$, 
where $Q_0^{\rm (Nb_3 Sn)}=1\times 10^{11}$ and $Q_0^{\rm (Nb)}=2\times 10^{10}$ are assumed.  
   }\label{fig9}
\end{figure*}

As seen in the last section, 
the layered Nb structure with $\lambda_1 > \lambda_2$ has a number of good properties. 
We can generalize it to materials other than Nb. 
One of the promising structures consists of a bulk clean Nb substrate with a penetration depth $\lambda_2$ and a thin dirty high-$B_c$ material with a penetration depth $\lambda_1 \, (> \lambda_2)$ and thickness $d \, ( < \lambda_1)$ (see the SS and SIS structures in Fig.~\ref{fig8}). 
\begin{enumerate}
\item Use of the thin dirty material at the surface and the clean bulk Nb as the substrate allows us to avoid the significant gap reduction effect at a high field and to retain a large $B_{c1}\simeq 170\,{\rm mT}$~\cite{Lin_Gurevich, Gurevich_review} (see Fig.~\ref{fig4} and Section.~\ref{section:DOS} again). 
\item The material combination that satisfies $\lambda_1 > \lambda_2$ leads to the reduction of the surface current and enhance the theoretical field limit~\cite{Kubo_APL, Gurevich_AIP, Posen_PhysRev} (see Section~\ref{section:field_limit} again). 
Fig.~\ref{fig9}(a) shows the theoretical field limit of the SS or SIS structures that consists of surface ${\rm Nb_3 Sn}$ layer and Nb substrate, where the insulator thickness is assumed to be negligible. 
It should be emphasized that this is nothing but a theoretical field limit for the ideal case. 
For a realistic case, there exist surface defects, and the practically achievable field becomes much different from Fig.~\ref{fig9}(a). 
For example, consider the case $d \gtrsim \lambda_1$, 
which is nothing but a bulk ${\rm Nb_3 Sn}$. 
Then the achievable field is given  by $B_0 \sim B_{c1}^{\rm (Nb_3 Sn)}$ instead of $B_{s}^{\rm (Nb_3 Sn)}$. 
\item The most important property equipped in the layered structures is the gimmick to stop vortex at an early stage of penetration in order for avoiding the vortex avalanches. 
In the SS structure, the SS boundary play the role of barrier to expel vortices if $\lambda_1 > \lambda_2$~\cite{Kubo_LINAC2014, Kubo_SUST, Checchin}, as mentioned in the last section (see Fig.~\ref{fig6} and Section~\ref{section:additional_barrier} again). 
In the SIS structure~\cite{Gurevich_APL}, a much different gimmick is available. 
Vortices disappear in the insulator layer and do not develop into avalanches. 
In addition, since vortex lines disappear in the insulator layer, 
only a small segment of vortex that pierce the surface layer can contribute to dissipation:  vortex dissipation is expected to be significantly suppressed~\cite{Gurevich_AIP, Kubo_SUST}. 
\item The surface resistance is suppressed because a part of surface current flows in the higher $B_c \,(\propto \Delta)$ material at the surface~\cite{Gurevich_AIP, Kubo_SUST} (see Section~\ref{section:Rs} again). 
Fig.~\ref{fig9}(b) shows $Q_0$ value of SS or SIS structure that consists of a surface ${\rm Nb_3 Sn}$ layer and Nb substrate, 
where the insulator thickness is assumed to be negligible. 
The horizontal axis represents as the ${\rm Nb_3 Sn}$ layer thickness $d$. 
\end{enumerate}
Note here the theoretical field limit for the ideal case becomes maximum when $d\simeq \lambda_1$ in Fig.~\ref{fig9}(a), 
but the practically achievable field does not necessarily become maximum at the same $d\simeq \lambda_1$. 
This can be understood by considering the following example. 
Suppose we have a ${\rm Nb_3 Sn}$ coated bulk Nb cavity with $d\simeq \lambda_1$ and there exists a material or topographic defect at the surface, 
leading to vortex penetration at a relatively low field (e. g., $B_0 \simeq 50\,{\rm mT}$). 
If the SS boundary barrier is so strong that vortices entering from the defect cannot go beyond the SS boundary, 
vortex dissipation is determined by its dynamics at $0\le x < d$. 
For the case this dissipation is so large that the cavity leads to a quench, 
we must reduce the thickness $d$ to improve the achievable field. 
The optimum thickness for maximizing the practically achievable field depends on the quality of material.

\subsection{What we see in experiments: \\
muon spin rotation, magnetization measurements, and vertical tests}\label{section:SSexperiments}

\begin{figure*}[tb]
   \begin{center}
   \includegraphics[width=0.9\linewidth]{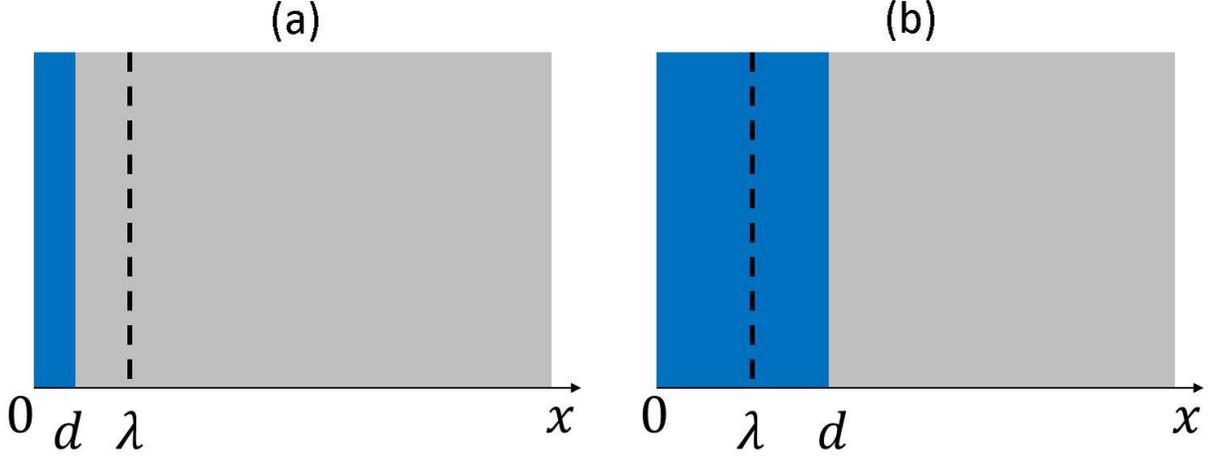}
   \end{center}\vspace{0cm}
   \caption{
SS structure with thickness $d$ smaller (a) and larger (b) than the penetration depth. 
   }\label{fig10}
\end{figure*}

As I repeatedly emphasized so far, 
the existence of the additional gimmick that pushes up the onset of vortex penetration is more important rather than the enhancement of the theoretical field limit. 
Figs.~\ref{fig5} and \ref{fig9}(a) show nothing but theoretical field limits for the ideal cases that come from the surface current reduction effect of the SS and SIS and do not represent the effects of the insulator layer or the SS boundary barrier,  
which become significant when the surface material is not ideal. 
To compare the theory with experiments, 
we need to carefully examine non-ideal cases.  
We assume there exist a lot of material or topographic defects at the surface (the BL surface barrier is not perfect). 
Furthermore, we assume the surface material has a much smaller $B_{c1}$ than Nb. 
We examine representative cases: 
$d\ll \lambda_1$ and $d\gg \lambda_1$ (see Fig.~\ref{fig10}). 
In the following, we focus on the SS structure.  
As for the SIS structure, see Refs.~\cite{Gurevich_AIP, Kubo_SUST}. 

%

\subsubsection{$d\ll \lambda_1$}

Let us begin with a simplified case: 
the barrier due to the SS boundary is so robust and a vortex that penetrate from the surface defect is always stopped by the barrier and does not go beyond the SS boundary. 
In this case, vortex penetration into the region $x>d$ occurs only when vortex rings are nucleated at $x=d$ due to a strong current density. 
This happens when $B_0 \sim B_s^{\rm (sub)}$, 
where the magnetic field attenuation in the surface layer is neglected because of the assumption $d\ll \lambda_1$. 
In $\mu$SR with a deep implantation ($\mu$ at $x \gg \lambda_1$) or magnetization measurements, 
we will see $B_{vp}=B_s^{\rm (sub)}$ (e.g., $B_s^{\rm (sub)}=B_s^{\rm (Nb)}\simeq 240\,{\rm mT}$ for Nb substrate). 
In vertical tests, cavities would withstand up to $B_s^{\rm (sub)}$. 
While vortices may enter from the surface defect at $B_0 < B_s^{\rm (sub)}$, 
they are stopped at $x=d\ll \lambda_1$ and vortex dissipation is expected to be negligible. 
For $B_0 > B_s^{\rm (sub)}$, vortices nucleated at $x=d$ flood into the material, 
and a cavity necessarily leads to a quench.

Next we consider the case the SS boundary barrier is moderately strong where vortices that enter from the surface defects can overcome the barrier and enter $x >d$ at $B_{c1}^{\rm (sub)} < B_0 < B_s^{\rm (sub)}$. 
Then, in $\mu$SR with a deep implantation and magnetization measurements, 
we will see $B_{c1}^{\rm (sub)} < B_{vp} < B_s^{\rm (sub)}$. 
In vertical tests, we will see $B_{c1}^{\rm (sub)} < B_{\rm quench} < B_s^{\rm (sub)}$ due to strong vortex dissipation that originate from vortices overcoming the SS boundary and moving at $0\le x < \lambda$. 
According to the $\mu$SR experiment~\cite{Laxdal_TTC2016, Laxdal_LCWS2016, Junginger_TTC2017},
$B_{vp}$ of Nb sample is given by $176\,{\rm mT}$ and $187\,{\rm mT}$ before and after $120^{\circ}$ baking, respectively: the dirty thin layer pushes up the onset of vortex penetration by $10\,{\rm mT}$. 
This is consistent with the above consideration where the interface between the dirty layer and the bulk clean Nb works as a moderately strong SS boundary barrier. 
Vertical test results are summarized in Ref.~\cite{Checchin}, 
where they also conclude baked Nb cavities pushes the onset of vortex penetration from $B_{c1}^{\rm (Nb)}$ to a field larger than $B_{c1}^{\rm (Nb)}$. 
Furthermore, $\mu$SR experiment results for ${\rm MgB_2}$ coated bulk Nb samples are also available~\cite{Laxdal_TTC2016, Laxdal_LCWS2016, Junginger_TTC2017}. 
Among their samples, 
the $d = 50\,{\rm nm}$ sample approximately corresponds with the present case ($d < \lambda_1$), 
which shows $B_{vp}\sim 220\,{\rm mT}$ and is also consistent with the above consideration.

As we increase $d$, the magnetic field and current density at $x=d$ are reduced because of the screening effect of the surface thin layer. 
Thus vortex nucleation at $x=d$ need $B_0 =c B_s^{\rm (sub)}$, 
where $c=B(0)/B(d) >1$ is calculable if the penetration depth of this sample is known. 
Then we have $B_{vp} \simeq c B_s^{\rm (sub)} > B_s^{\rm (sub)}$ for the case the SS boundary barrier is perfectly robust. 
For a moderately strong SS boundary barrier where vortices in the surface material can overcome the SS boundary barrier, 
we have $B_{c1}^{\rm (sub)} < B_{vp} <  c B_s^{\rm (sub)}$. 
According to the magnetization measurements of ${\rm MgB_2}$ coated bulk Nb sample~\cite{Tan_SR}, 
$B_{vp}\simeq 260\,{\rm mT}$ at $T=2.8\,{\rm K}$ for $d=200\,{\rm nm}$. 
In addition, $\mu$SR experiment results of ${\rm MgB_2}$ coated bulk Nb samples with thickness $d=150\,{\rm nm}$~\cite{Laxdal_TTC2016, Laxdal_LCWS2016, Junginger_TTC2017} shows $B_{vp}\sim 230\,{\rm mT}$. 
These results are also consistent with the above consideration and demonstrate the SS boundary barrier pushes up the onset of vortex penetration.

As will be shown below, $d\gg \lambda_1$ yields $B_{\rm quench} \sim B_{c1}^{\rm (layer)}$ in SRF application. 
Exploring the optimum thickness in the range $0< d \lesssim \lambda_1$ would be the next important task for SRF application.

\subsubsection{$d\gg \lambda_1$}

In common with the above, we start from the case the SS boundary barrier is perfectly robust: vortices that penetrate from the surface defects are always stopped by the barrier and does not go beyond the SS boundary. 
Since $d\gg \lambda_1$, the surface layer can be regarded as a bulk superconductor. 
Vortices will start to penetrate the surface material at $B_0 \sim B_{c1}^{\rm (layer)}$ (strictly, $B_{c1}^{\rm (layer)} < B_0 < B_{s}^{\rm (layer)}$) (e.g., $B_{c1}^{\rm (layer)} = B_{c1}^{\rm (Nb_3 Sn)} \simeq 50\,{\rm mT}$ for thick ${\rm Nb_3 Sn}$ coated Nb). 
These vortices, however, are stopped by the SS boundary barrier at $x=d$, 
and the magnetic flux is absent at $x>d$ at this stage. 
Thus, in $\mu$SR experiments, 
we cannot see any signal if the $\mu$ implantation depth is much larger than $d$, 
even though vortices are already inside the surface layer.   
In vertical tests, on the other hand, cavities will suffer quenches at this stage,  
$B_{\rm quench} \sim B_{c1}^{\rm (layer)}$, 
because strong vortex dissipation starts at $B_0\sim B_{c1}^{\rm (layer)}$. 
Note that the $\mu$SR with the implantation depth $\lesssim d$ can confirm the fact that vortices start to penetrate at $B \sim B_{c1}^{\rm (layer)}$.

In the $\mu$SR experiment, we can further increase $B_0$. 
While the magnetic flux in the surface layer increases, 
the magnetic flux at $x>d$ does not change, 
because vortices in the surface layer cannot go beyond $x= d$. 
The magnetic flux at $x>d$ starts to increase when vortex nucleation begins at $x=d$. 
The nucleation is triggered when the current density at $x=d$ exceeds the depairing current (e. g., $J \sim B_s^{\rm (Nb)}/\mu_0 \lambda_{\rm Nb}$ for Nb substrate). 
This happens the magnetic flux density in the surface material $\bar{B}$ achieves $\bar{B} \sim B_s^{\rm (sub)}$.  
Since $\bar{B}$ is approximately given by $B_0$ when $B_0 \gg B_{c1}^{\rm (layer)}$, 
we find the vortex ring nucleation at $x=d$ start at $B_0 \sim B_s^{\rm (sub)}$. 
Thus, in $\mu$SR experiments (with $\mu$ much deeper than $d$), 
we will see $B_{vp} \sim B_s^{\rm (sub)}$, 
which is the similar value as the case $d \ll \lambda_1$. 
Note here this $B_{vp}$ for the case $d\gg \lambda_1$ is not related to the SRF field limit, 
because vortex penetration starts at $B_0 \sim B_{c1}^{\rm (layer)}$ and a cavity suffers a quench due to a strong vortex dissipation: 
$B_{\rm quench} \sim B_{c1}^{\rm (layer)}$.

According to $\mu$SR experiments~\cite{Laxdal_TTC2016, Laxdal_LCWS2016, Junginger_TTC2017}, 
${\rm Nb_3 Sn}$ ($d=2\,{\rm \mu m}$) coated bulk Nb sample shows $B_{vp}\sim 220\,{\rm mT}$, 
and ${\rm Mg B_2}$ ($d=300\,{\rm nm}$) coated bulk Nb sample shows $B_{vp}\sim 230\,{\rm mT}$. 
In addition, there are a lot of vertical test results for thick ${\rm Nb_3 Sn}$ coated bulk Nb cavity, 
where $B_{\rm quench} \sim B_{c1}^{\rm (Nb_3 Sn)}$~\cite{Posen_SUST}. 
These all are consistent with the above consideration. 
It should be emphasized again that $B_{vp} \gg B_{c1}^{\rm (layer)}$ for $d\gg \lambda_1$ is not related to the SRF field limit as mentioned above, 
but its value much larger than $B_{c1}^{\rm (layer)}$ is quite meaningful: 
it indeed demonstrates the SS boundary works as a barrier.

Now we consider the case the SS boundary barrier is so weak that vortices in the surface material can overcome the barrier and go into $x>d$. 
The vertical test results will not change: $B_{\rm quench} \sim B_{c1}^{\rm (layer)}$. 
This is because quenches are triggered by vortex penetration at $B_0 \sim B_{c1}^{\rm (layer)}$ and the SS boundary at $x=d \gg \lambda_1$ does not play any role. 
The $\mu$SR measurements results will become $B_{vp} < B_s^{\rm (sub)}$,
because vortices in the surface material can overcome the SS boundary barrier and the flux at $x>d$ start to increase even when $B_0 < B_s^{\rm (sub)}$ as long as the barrier is weak enough. 
The concrete value of $B_{vp}$ depends on the strength of the SS boundary barrier.

Finally, it should be noted that all the above experiments are consistent with the existence of the SS boundary barrier, 
but other mechanisms can also explain the experimental results (e. g., strong pinning centers may pin vortices at the SS boundary).  
We do not know which mechanism is dominant. 
To reveal how the SS boundary prevents vortex penetration is also the next task.

\subsubsection{SIS}

For the SIS case, 
vortices that penetrate from the surface defect vanish at the insulator layer. 
This is robust gimmick to stop vortex penetration rather than the SS boundary of the SS structure. 
Furthermore, the vortex dissipation comes from the small segments that penetrate the surface layer and is expected to be significantly reduced~\cite{Gurevich_AIP}. 
See also Ref.~\cite{Kubo_SUST} for a review of SIS, 
where theoretical aspects are explained in so detail that readers can follow all the calculations.

\section{Summary}

We briefly reviewed basics of SRF (Section~\ref{section:basics}) and looked back the history (Section~\ref{section:history}). 
Use of pure Nb was not enough to achieve $E_{acc} > 40\,{\rm MV/m}$. 
The low temperature baking was the key for high gradients. 
The recent experimental results on the low temperature baking was also introduced:  
the baking changes a simple clean Nb to a layered structure that consists of a thin dirty Nb and clean bulk Nb. 
We then summarized the multiple benefit of the layered Nb structure realized in the baked Nb (Section~\ref{section:benefit}): 
(1) the gap of dirty superconductor is well behaved at a high field rather than clean superconductor; 
(2) suppress the surface current and enhance the theoretical field limit; 
(3) prevent the vortex penetration by the additional barrier; 
(4) reduce surface resistance if the surface layer has a lower surface resistance. 
The multiple benefit of the layered Nb structure was generalized to materials other than Nb, and next SRF technologies, SS and SIS structures, were introduced  (Section~\ref{section:next}). 
These are expected to lead to much higher gradients than the present technologies.
Here we emphasized that implementation of a gimmick to stop vortex penetration is essential in order for going beyond $B_{c1}$.  
We then explained SS structure in detail and focusing on what we see in experiments if the surface material is not ideal one (Section~\ref{section:SSexperiments}). 
Experimental data seem to support the validity of SS structure and demonstrate the SS boundary works as a barrier. 
To explore the optimum thickness of the surface layer ($d\le \lambda_1$) would be the next task. 
However, other mechanism (e.g., strong pinning centers) can also explain the experimental results. 
To reveal how the SS boundary stop vortices is also the next task. 
Detailed discussion on the SIS structure is given in Refs.~\cite{Gurevich_AIP, Kubo_SUST}.

\begin{acknowledgments}

The present work is supported by JSPS Grant-in-Aid for Young Scientists (B) 26800157, 
JSPS Grant-in-Aid for Challenging Exploratory Research 26600142, 
and Photon and Quantum Basic Research Coordinated Development Program from the Ministry of Education, Culture, Sports, Science and Technology, Japan. 

\end{acknowledgments}

\bibliography{basename of .bib file}

\begin{thebibliography}{99}

\bibitem{Grassellino_N_dope}
A. Grassellino et al, 
Supercond. Sci. Technol. {\bf 26}, 102001 (2013). 

\bibitem{Dhakal_Ti_dope_PRAB}
P. Dhakal et al., 
Phys. Rev. ST Accel. Beams {\bf 16}, 042001 (2013). 

\bibitem{Dhakal_Ti_dope_IPAC2012}
P. Dhakal et al., 
in proceedings of IPAC2012, New Orleans, Louisiana, USA (2012), p. 2426, WEPPC091.




\bibitem{Ciovati_IPAC2013}
G. Ciovati, 
in {\it Proceedings of IPAC2013, Shanghai, China} (2013), p. 3124, THYB201. 

\bibitem{saitoEP}
K. Saito et al., 
in {\it Proceedings of SRF1989, KEK, Tsukuba, Japan} (1989), p. 635, SRF89G18. 

\bibitem{furuyaEP}
T. Furuya, 
in {\it Proceedings of SRF1989, KEK, Tsukuba, Japan} (1989), p. 305, SRF89D02. 

\bibitem{bernerdHPR}
Ph. Bernard, D. Bloess, T. Flynn, C. Hauviller, W. Weingarten, P. Bosland, and J. Martignac, 
in {\it Proceedings of EPAC1992, Berlin, Germany} (1992), p. 1269. 

\bibitem{saitoHPR}
K. Saito, H. Miwa, K. Kurosawa, P. Kneisel, S. Noguchi, E. Kako, M. Ono, T. Shishido and T. Suzuki, 
in {\it Proceedings of SRF1993, CEBAF, Newport News, Virginia, USA} (1993), p. 1151, SRF93J03. 

\bibitem{kneiselHPR}
P. Kneisel, B. Lewis and L. Turlington, 
in {\it Proceedings of SRF1993, CEBAF, Newport News, Virginia, USA} (1993), p. 628, SRF93I09. 

\bibitem{kojimaCA}
Y. Kojima et al., 
in {\it Proceedings of SRF1989, KEK, Tsukuba, Japan} (1989), p. 85, SRF89A07. 

\bibitem{kako_bake}
E. Kako et al., 
in {\it Proceedings of SRF1995, Gif-sur-Yvette, France} (1995), p. 425, SRF95C12. 

\bibitem{kneisel_40}
P. Kneisel, R. W. R${\rm \ddot{o}}$th and H. - G. Kiirschner,  
in {\it Proceedings of SRF1995, Gif-sur-Yvette, France} (1995), p. 449, SRF95C17. 

\bibitem{ono_bake}
M. Ono et al., 
in {\it Proceedings of SRF1997, Abano Terme (Padova), Italy} (1997), p. 472, SRF97C08. 

\bibitem{lilje_bake}
L. Lilje et al., 
in {\it Proceedings of SRF1999, La Fonda Hotel, Santa Fe, New Mexico, USA} (1999), p. 74, TUA001. 


\bibitem{Romanenko_bake}
A. Romanenko et al., 
Appl. Phys. Lett. {\bf 104}, 072601 (2014). 





\bibitem{Lin_Gurevich}
F. P. Lin and A. Gurevich, 
Phys. Rev. B {\bf 85}, 054513 (2012). 

\bibitem{Gurevich_review}
A. Gurevich, 
Rev. Accel. Sci. Technol. {\bf 5}, 119 (2012). 


\bibitem{Kubo_APL}
T. Kubo, Y. Iwashita, and T. Saeki, 
Appl. Phys. Lett. {\bf 104}, 032603 (2014).  

\bibitem{Kubo_LINAC2014}
T. Kubo, 
in {\it Proceedings of LINAC14, Geneva, Switzerland} (2014), p. 1026, THPP074. 

\bibitem{Gurevich_AIP}
A. Gurevich, 
AIP Advance {\bf 5}, 017112 (2015). 

\bibitem{Posen_PhysRev}
S. Posen, M. K. Transtrum, G. Catelani, M. U. Liepe, and J. P. Sethna, 
Phys. Rev. Applied {\bf 4}, 044019 (2015). 

\bibitem{Kubo_SUST}
T. Kubo, 
Supercond. Sci. Technol. {\bf 30}, 023001 (2017). 

\bibitem{Checchin}
M. Checchin, A. Grassellino, M. Martinello, S. Posen, A. Romanenko, and J. F. Zasadzinski, 
in {\it Proceedings of IPAC2016, Busan, Korea} (2016), p. 2254, WEPMR002. 


\bibitem{Grassellino_bake}
A. Grassellino and S. Aderhold,  
``New Low T Nitrogen Treatments Cavity Results with Record Gradients and Q", 
TESLA Technology Collaboration (TTC) meeting, Saclay, France (2016). 

\bibitem{Grassellino_bake_arxiv}
A. Grassellino et al., 
``Unprecedented Quality Factors at Accelerating Gradients up to 45 MV/m in Niobium Superconducting Resonators via Low Temperature Nitrogen Infusion", 
arXiv:1701.06077 [physics.acc-ph].


\bibitem{Gurevich_APL}
A. Gurevich, 
Appl. Phys. Lett. {\bf 88}, 012511 (2006).  


\bibitem{Laxdal_TTC2016}
R. Laxdal et al., 
``New insights for reaching higher gradients from muSR samples studies", 
TESLA Technology Collaboration (TTC) meeting, Saclay, France (2016). 

\bibitem{Laxdal_LCWS2016}
R. Laxdal, 
``Report from TRIUMF", 
The International Workshop on Future Linear Colliders, LCWS2016, Morioka, Japan (2016). 


\bibitem{Junginger_TTC2017}
T. Junginger, W. Wasserman, and R. Laxdal, 
``Insights into multilayers and new materials from muSR", 
TESLA Technology Collaboration (TTC) meeting, Michigan, USA (2017). 

\bibitem{Tan_SR}
T. Tan, M. A. Wolak, X. X. Xi, T. Tajima, and L. Civale, 
Scientific Reports {\bf 6}, 35879 (2016).  

\bibitem{Posen_SUST}
S. Posen and D. L. Hall, 
Supercond. Sci. Technol. {\bf 30}, 033004 (2017). 

\end{thebibliography}

\end{document}